\documentstyle[12pt]{article}
\begin{document}
\title{Variational Wave Function for Generalized Wigner Lattices in
  One Dimension}
\author{Simone Fratini$^1$, Belen Valenzuela$^2$ and Dionys Baeriswyl$^3$}
\date{}
\maketitle
\begin{enumerate}
\item[   $^1$]  Laboratoire d'Etudes des Propri\'et\'es Electroniques
  des Solides,  CNRS, \\ 25 avenue des Martyrs, B.P.\ 166, 
F-38042 Grenoble Cedex 9, France
\item[   $^2$] Instituto de Ciencia de Materiales, CSIC, Cantoblanco\\
E-28049 Madrid, Spain
\item[   $^3$] Department of Physics, University of Fribourg, P\'erolles\\
CH-1700 Fribourg, Switzerland
\end{enumerate}

\begin{abstract}
We study a system of electrons  on a
one--dimensional lattice,
interacting through the long range Coulomb forces, by means of a variational
technique which is the strong coupling analog of the Gutzwiller
approach.  The problem is thus the quantum version of Hubbard's classical
model of the generalized Wigner crystal 
[J. Hubbard, Phys. Rev. B {\bf 17}, 494 (1978)]. The magnetic
exchange energy arising from quantum fluctuations is calculated, and
turns out to be smaller than the energy scale governing charge degrees
of freedom. This approach could be relevant in insulating
quasi--one--dimensional compounds where the long range
Coulomb interactions are not screened. In these compounds charge order
often appears at high temperatures and coexists with magnetic order at
low temperatures.
\end{abstract}
\thispagestyle{empty}
\pagestyle{empty}

\section{INTRODUCTION}
The Luttinger Liquid is a paradigm for models of one--dimensional
interacting electrons. Remarkably, it not only holds for weak (bare) couplings,
but remains valid up to strong couplings (for the special case of
the Hubbard model up to $U=\infty$). For short--range interactions, a small
set
of coupling constants, corresponding to forward,
backward and Umklapp scattering, determines the low--energy behavior.
For long--range interactions, such as the
unscreened Coulomb interaction, the dependence of the forward
scattering on momentum transfer $q$ has to be taken into account, as this term
diverges logarithmically for $q\rightarrow 0$. Nevertheless, the method of
bosonization, which is so useful for weak short--range couplings, can be
extended to this case. As shown by Schulz, this method predicts
a ground state with quasi--long--range charge order for the homogeneous
electron
gas with long--range ($1/r$) Coulomb interaction \cite{schulz}. This behavior
is found to hold up to the limit of a very dilute gas where the Coulomb
interaction dominates and the ground state is a (Wigner) crystal of electrons
with strongly localized wave functions.

For narrow--band materials the effects of the underlying lattice have to be
taken into account. For the case of Coulomb interactions this has been
pointed out by Hubbard \cite{hubbard}, who considered the extreme limit of
zero bandwidth. In this case the problem is equivalent to that of a system of
classical charges distributed over the sites of a lattice and coupled to
each other by the Coulomb interaction. We have extended Hubbard's
considerations to a more realistic model including a small but finite hopping
term. Using a variational wave function
we are not only able to describe the incipient charge delocalization, but
also to account for an antiferromagnetic interaction induced by the exchange
of electrons located on the neighboring sites of the Wigner lattice
\cite{valenzuela}. For a small
hopping amplitude $t$ the energy scale for the charge degrees of freedom
turns out to be much larger than that for the spin degrees of
freedom.

\section{VARIATIONAL APPROACH}
We consider a system of one--dimensional fermions interacting via a
local repulsion $U$ and a long--range Coulomb potential $V_m=V/|m|$. The
corresponding tight--binding Hamiltonian is
\begin{equation}
H= -t \sum_{i\sigma} \left( c_{i,\sigma}^+ c_{i+1,\sigma} + h.c.\right)
+ U \sum_i  n_{i\uparrow}n_{i\downarrow}
+\frac{1}{2}\sum_{i\neq j} V_{i-j} n_i n_j\ ,
\label{hamiltonian}
\end{equation}
where $n_i = n_{i\uparrow}+n_{i\downarrow}$ measures the density per site.
The Fourier transform of the Coulomb potential is $V(q)\sim-2V{\log} q$
for $q\rightarrow 0$ (we set the lattice constant equal to 1). In the
following we limit ourselves to the case of a quarter--filled band
(i.e., an average electron density of $n=1/2$ per site). Hubbard's
classical solution is then an alternation of occupied and empty
sites, corresponding to a $4k_F$ charge modulation and uncoupled spin
degrees of freedom.

For finite but small $t$ we use the variational ansatz
\begin{equation}
  |\Psi_B\rangle= e^{-\eta \hat{T}}|\Psi_\infty\rangle\ ,
  \label{ansatz}
\end{equation}
where $\eta$ is a variational parameter,  $\hat{T}$ is the kinetic energy
operator and $|\Psi_\infty\rangle$ is the ground state for $t=0$,
i.e., Hubbard's classical solution. This wave function, introduced as
a counterpart of the Gutzwiller ansatz to describe the ground state of
the large $U$ Hubbard model at half filling \cite{dionys}, has been
successfully applied to the Mott--Hubbard transition \cite{dzierzawa}.
The role of the operator $e^{-\eta \hat{T}}$ is analogous to
that of $e^{-\lambda \hat{D}}$ in Gutzwiller's wavefunction, where double
occupancy
is suppressed in order to reduce the weight of configurations with high
potential energy. Here the factor $e^{-\eta \hat{T}}$ suppresses
states with high kinetic energy.
In the limit $\eta\rightarrow \infty$ only the state with the
lowest kinetic energy, i.e., the Fermi sea, survives.

For very large $U$, where double occupancy is expected to be completely
suppressed, the charge degrees of freedom can be described in terms of
spinless fermions coupled by long--range Coulomb forces, whereas the spin
degrees of freedom are uncoupled. For spinless fermions, the variational
energy is readily calculated and turns out to be very close to the
Hartree--Fock approximation \cite{valenzuela}. The relevant scale
for charge
excitations is the energy required for moving an electron from its position
in the classical Hubbard configuration to an empty neighboring site, i.e.
\begin{equation}
\Delta_c= V\sum_{l=1}^{\infty}\frac{1}{l[(2l)^2-1]}\approx 0.39 V\ .
\label{chargegap}
\end{equation}
Note that the charge
gap would be larger ($\Delta_{nn}=0.5V$) if we only retained
interactions between electrons on nearest neighbors. This simple argument
shows that even though the ground state configuration  is well
described by an ``extended'' Hubbard model, the spectrum of excited states is
different when the long-range tail of the Coulomb potential is taken
into account. In the latter case, the system is ``softer'' with
respect to charge fluctuations.

For finite but large values of $U$, the double occupancy does not vanish
but is expected to be small. In order to take this effect into account we
consider a refined wave function
\begin{equation}
|\Psi_{BG}\rangle=
e^{-\lambda \hat{D}}e^{-\eta \hat{T}}|\Psi_\infty\rangle\ ,
\label{ansatz1}
\end{equation}
where $\hat{D}$ measures the number of doubly occupied sites.
The starting ground state $|\Psi_\infty\rangle$ is now a superposition of
all possible configurations of spins attached to the even
(or odd) sites of the chain. The operator $e^{-\eta \hat{T}}$ again
controls the delocalization of electrons away from their classical site,
while the Gutzwiller operator $e^{-\lambda \hat{D}}$ reduces the weight of
configurations with doubly occupied sites.
The ansatz (\ref{ansatz1}), introduced for the Hubbard model by Otsuka
\cite{otsuka}, leads to a dramatic improvement of the ground state energy,
as demonstrated for the exactly soluble one--dimensional Hubbard model
with long--range hopping \cite{dzierzawa2}.
For small $t$, the energy can
be worked out as an expansion in $t/V$, in close analogy to the
procedure used for the Hubbard model \cite{dionys}. In the large $U$ limit we
obtain for the minimum of the variational energy per site \cite{valenzuela}
\begin{equation}
\epsilon_{BG}= -\frac{1}{2{\log}2-1}\frac{t^2}{V} +
J\ \langle\Psi_{\infty}|
\sum_i\ '\left({\bf S}_i{\bf S}_{i+2}-\frac{1}{4}n_in_{i+2}\right)
|\Psi_{\infty}\rangle\ ,
\label{energy}
\end{equation}
where the sum runs over all even sites and the exchange constant is given by
\begin{equation}
J=\frac{36{\log}2\ t^4}{(15-16{\log}2)(2{\log}2-1)^2V^3+6(2{\log}2-1)^2V^2U}\ .
\end{equation}
The first term in Eq.\ (\ref{energy}) is identical to the variational energy
for spinless electrons in the large $V$ limit (or the spinful case for
$U\rightarrow \infty$). Here we see that for finite $U$ we obtain an
antiferromagnetic coupling between the spins. The
remaining problem of finding the best magnetic state is equivalent to the
problem of determining the ground state of the one--dimensional Heisenberg
model. Its solution is known thanks to Bethe. Recently it has been
shown that the spin correlations decay like
$(-1)^{(i-j)}(${log}$|i-j|)^{\frac{1}{2}}/|i-j|$ \cite{auerbach}.
Therefore our variational wave function exhibits long--range charge order
and algebraic magnetic order.

\section{DISCUSSION}

In this work, we have studied a one-dimensional system of electrons
interacting through the long-range Coulomb forces. Starting from the
strong coupling generalized Wigner lattice, we have introduced a variational
wave function which allows to treat the effects of quantum
fluctuations. As a result, magnetic correlations develop out of the
charge ordered configuration, with a lower energy scale.

Quasi one-dimensional organic compounds of
the (TMTTF)$_2$X family  \cite{nad}, as well as the inorganic materials
(DI-DCNQI)$_2$X \cite{hiraki},
are known to exhibit charge ordered structures
at temperatures $T\sim 100-200K$.
There are several experimental indications that the long-range
electron-electron interactions are the common driving mechanism
of the charge ordering:
(i) the measured electronic conductivities are low, suggesting
that the long-range tail of the Coulomb potential is not screened;
(ii) a strong $4k_F$ superstructure, inferred from both X-ray and NMR
spectroscopy,
develops in the charge ordered region,
which can not be ascribed to an ordinary $2k_F$ Fermi surface
instability; (iii) the $4k_F$ ordering is not necessarily
associated to a structural transition;
(iv) charge ordering sets in at a much higher temperature than
magnetic ordering.

In such compounds, where the filling is fixed by stoichiometry to
one carrier every two sites, electron-electron correlations  are generally
treated theoretically in the framework of the  extended Hubbard model, which
only retains on-site and nearest--neighbors interactions \cite{exthub}.
Although this
can successfully reproduce the $4k_F$ charge correlations,
new physics can in principle be
expected if the full long--range potential is taken
into account. For example, in the metallic regime, it is known that
quasi--long--range order appears in purely one--dimensional systems
due to the strong forward scattering
associated to the $1/r$ behavior at long distances, regardless
of the interaction strength \cite{schulz}.
In the charge ordered regime, which is the object of
the present work,  the spectrum of excited states can differ
substantially from what expected in the extended Hubbard model. The
charge gap is lower, leading to stronger charge fluctuations, and to
larger magnetic exchange energies.
In compounds where the filling differs from
$n = 1/2$, such as the TTF TCNQ (studied by Hubbard \cite{hubbard}),
the use of the extended Hubbard model
is even more debatable.

Clearly, the phenomenology observed in quasi--one--dimensional
compounds is much more
complex  than what emerges from the simple model considered
here. The detailed phase diagrams are determined by the chain
dimerization, anion
size and symmetry,   inter-chain couplings, etc. However, the
ubiquitous experimental
signatures of the electron-electron  interactions  in such
insulating  systems call for a more systematic study of the role of
the long-range Coulomb interactions.

\section*{Acknowledgments}
We wish to thank F. Nad and J.-P. Pouget for fruitful discussions.

\end{document}